\documentclass[runningheads]{llncs}
\usepackage[T1]{fontenc}
\usepackage{graphicx}
\usepackage{amsmath}
\usepackage{amssymb}
\usepackage{hyperref}
\usepackage{cleveref}
\usepackage{float}
\usepackage{siunitx}
\usepackage{tabularx}
\usepackage{booktabs}
\usepackage{multicol}
\usepackage{multirow}
\usepackage{adjustbox}
\usepackage{rotating}
\usepackage{verbatim}
\usepackage{array}
\usepackage{multirow}
\usepackage{notoccite}
\newcolumntype{P}[1]{>{\centering\arraybackslash}p{#1}}
\begin{document}
\title{Groupwise Registration with Physics-Informed Test-Time Adaptation on Multi-parametric Cardiac MRI}
\titlerunning{Groupwise Registration with Physics-Informed Test-Time Adaptation}

\author{Xinqi Li\inst{1,2,4}, Yi Zhang\inst{3}, Li-Ting Huang\inst{1}, Hsiao‐Huang Chang\inst{5}, Thoralf Niendorf\inst{2}, Min-Chi Ku\inst{2}, Qian Tao\inst{3}, Hsin-Jung Yang\inst{1}$^*$}  
\authorrunning{Xinqi Li et al.}
\institute{Biomedical Imaging Research Institute, Cedars-Sinai Medical Center, United States \and 
Berlin Ultrahigh Field Facility (B.U.F.F.), Max Delbrück Center for Molecular Medicine in the Helmholtz Association (MDC), Germany \and Department of Imaging Physics, Delft University of Technology, Netherlands \and
TUM School of Computation, Information and Technology, Technische Universität München, Germany \and
Department of Surgery, Taipei Veterans General Hospital, Taiwan}

\maketitle              
\begin{abstract}
Multiparametric mapping MRI has become a viable tool for myocardial tissue characterization. However, misalignment between multiparametric maps makes pixel-wise analysis challenging. To address this challenge, we developed a generalizable physics-informed deep-learning model using test-time adaptation to enable group image registration across contrast weighted images acquired from multiple physical models (e.g., a T1 mapping model and T2 mapping model). The physics-informed adaptation utilized the synthetic images from specific physics model as registration reference, allows for transductive learning for various tissue contrast. We validated the model in healthy volunteers with various MRI sequences, demonstrating its improvement for multi-modal registration with a wide range of image contrast variability.


\end{abstract}
\section{Introduction}
Multiparametric quantitative cardiac magnetic resonance (CMR) imaging has become an essential diagnostic tool for cardiovascular pathology. The myocardial multiparametric mapping techniques, including T1 mapping and T2 mapping, provide complementary information about the heart and can reflect various myocardial abnormalities that can be used as a crucial tool to differentiate subtypes of cardiomyopathies~\cite{sado2013identification}, and stage ischemic injuries~\cite{messroghli2016clinical,lopez2017multiparametric,wang2024multi,eichhorn2022multiparametric,hassani2020myocardial}. 
However, the cardiac and respiratory motion introduce susceptibility to image alignment. Such motion can substantially compromise the fitting of quantitative maps and the multiparametric analysis between different mapping sequences~\cite{xue2012motion,kellman2013t1}. Motion correction has therefore evolved into a fundamental component of contemporary post-processing workflows~\cite{ashburner2007fast,chen2021deep}, ensuring optimal image quality and diagnostic accuracy~\cite{makela2002review}. 

Deep learning-based image registration has gained significant attention in medical imaging due to its ability to provide rapid inference compared to traditional optimization-based methods~\cite{arava2021deep,balakrishnan2019voxelmorph,fu2020deep,hanania2023pcmc,li2021learning,li2023contrast,li2022motion,martin2020groupwise,yang2022end}, offering computational efficiency and generalizability to in-domain data~\cite{balakrishnan2019voxelmorph,fu2020deep}. However, these methods face several limitations, particularly in multiparametric MRI mapping. One key challenge is the large signal variations between weighted images, making them less robust in quantitative MRI scenarios. Additionally, existing physics-informed registration methods, while showing improvement by incorporating known MRI signal models~\cite{zhang2024deep}, are often restricted to a single physics model within a single mapping acquisition. This limitation reduces their effectiveness when registering multiparametric maps acquired under different MRI sequences, where signal evolution follows multiple distinct models. Conventional methods that adopt motion-free synthetic images with similar contrast to the targeted data for registration show promising results~\cite{xue2012motion}. However, training neural networks using synthetic images has two main limitations: it constrains the model's applicability to a specific modality with known signal model, while demanding substantial computational resources. Recent advances in large pre-trained models that learn good representations and provide strong foundations for fine-tuning, have renewed interest in test-time adaptation~\cite{sun2020test}. This allows model optimization based on existing models~\cite{li2023contrast} and specifically fine-tuning for the given task without redoing the intensive pre-training step, which sheds light on the registration between multiparametric maps. 

Here, we propose a new method for test-time adaptation of cardiac registration that pre-trains the neural network using a model-agnostic approach~\cite{li2023contrast} and leverages physical models of MR contrast evolution as additional guidance during the fine-tuning process. By fine-tuning the model at test time, it can adapt specifically to the contrast patterns present in the current image series, enabling more robust registration across diverse contrast mechanisms. Furthermore, cardiac motion patterns and tissue appearances can vary significantly between patients due to differences in heart rate, breathing patterns, and underlying pathologies. Test-time adaptation allows the model to optimize its parameters for each specific case, accounting for patient-specific variations without requiring a massive training dataset that covers all possible scenarios. This adaptive capability is especially valuable when dealing with pathological cases that may be underrepresented in training data. 

In this work, we proposed a registration pipeline that leverage the contrast agnostic pretraining and physics guided test-time adaptation to enable a generalizable registration technique for multiparametric CMR maps. Our main contributions are:
\begin{enumerate}
    \item Developed a generalizable deep-learning model to enable group image registration, regardless of contrast progression, length of the image series. 
    \item Combined an existing rPCA-based contrast agnostic registration pipeline with physics models to guide model fine-tuning and boost registration accuracy when large contrast variations (e.g., weighted images from T1 recovery and T2 decay) are present.
    \item Integrated a hierarchical registration pipeline to incorporate image contrast ruled by multiple physical models. 
\end{enumerate}

\section{Methods}

\subsection{Problem Formulation}
The multi-parametric dataset contains a set of image series $\mathcal{S} = \{S^1, S^2, \cdots, S^M\}$ where $M$ is the total number of sequences. Each sequence $S_i$ contains its time series images $S^i = \{I^i_1, I^i_2, \cdots, I^i_N\}$ where $i \in \{1, \cdots, M\}$ is the sequence index and $N$ is the number of time frames, $I^i_t \in \mathcal{R}^{H \times W}$ is the image at inversion time $t$ in sequence $i$. The goal of the registration is to achieve spatial alignment among all $I^i_t$ by determining each frame's corresponding deformation field $\phi^i_t \in \mathcal{R}^{2 \times H \times W}$. The deformable mapping can be obtained by solving the following optimization problem:
\begin{equation}
\phi^* = \mathop{\arg \min}\limits \mathcal{L}_\mathrm{similarity} + \lambda_0 \mathcal{L}_\mathrm{smooth} + \lambda_1 \mathcal{L}_\mathrm{cyclic}.
\label{eq_loss}
\end{equation}
The $\mathcal{L}_\mathrm{similarity}$ term measures the groupwise image similarity and will be introduced in detail in \Cref{sec:PITT}. The $\mathcal{L}_\mathrm{smooth}$ term denotes the smoothness regularization and is regularized by 
\begin{equation}
\mathcal{L}_{\text {smooth }}=\frac{1}{H \times W} \sum_{t=1}^{N}\int_0^H \int_0^W\left[\left(\frac{\partial^2 \phi_i^t}{\partial x^2}\right)^2+\left(\frac{\partial^2 \phi_i^t}{\partial y^2}\right)^2 +2\left(\frac{\partial^2 \phi_i^t}{\partial x y}\right)^2\right]d x d y.
\end{equation}
Furthermore, given that deformation fields in the group exhibit periodic or symmetric properties, $\mathcal{L}_\mathrm{cyclic}$ term denotes the cyclic consistency and is defined as follows:
\begin{equation}
\mathcal{L}_\mathrm{cyclic} = \sqrt{\frac{1}{2(H\times W)}\sum_{i, j \in H, W}\left(\sum_{t=1}^N \phi_i^t(i, j)\right)^2},
\label{eq2}
\end{equation}
to avoid the degenerate case in which textural features collapse across the entire image sequence~\cite{li2023contrast}. $\lambda_0$ and $\lambda_1$ denote the weight parameters respectively.
\begin{figure}[ht!]
    \centering
    \includegraphics[width=0.97\textwidth]{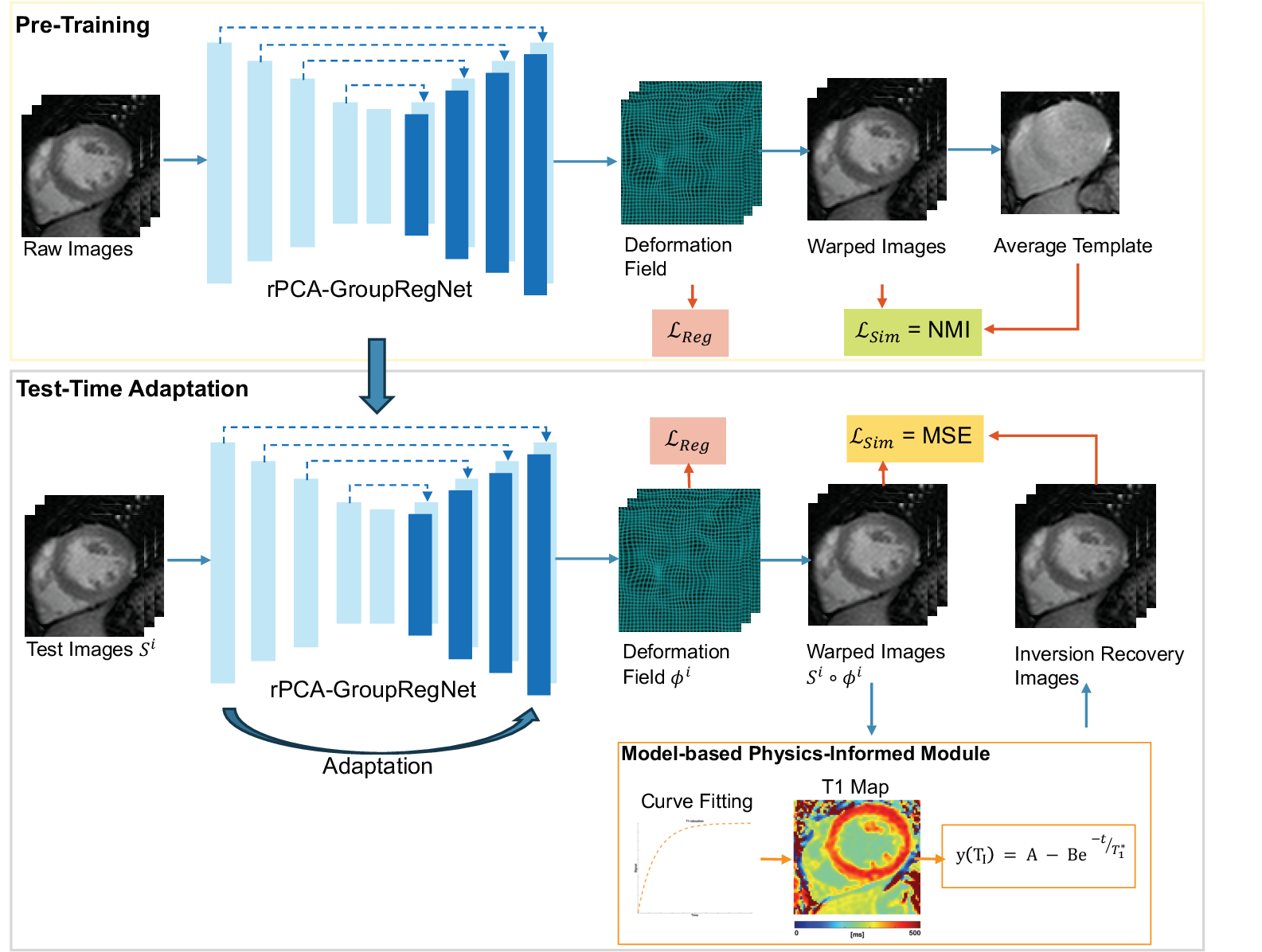}
    \caption{Overview of the physics-informed test-time adaptation. The model is pre-trained using training data, and the pre-trained model is adapted to obtain desired capabilities on specific test images according to different physics models. The similarity term is changed from normalized mutual information to mean squared error during fine-tuning. In this figure, we use the data following the $T_1$-relaxation model as an example. In practice, it can be replaced with others based on the data modality. The rPCA-GroupRegNet is built based on the previous work \cite{li2023contrast}}
    \label{fig:tta}
\end{figure}
\subsection{Physics-Informed Test-time Adaptation}
\label{sec:PITT}
Test-time adaptation is a special setting of unsupervised domain adaptation where a trained model on the source domain has to adapt to the target domain~\cite{9880363}. Test-time fine-tuning is one of the approaches that works by fine-tuning the parameters of a pre-trained model at test time~\cite{hardt2024test}. The overview of the proposed approach is shown in \Cref{fig:tta}. 

In pre-training, an average template $S_{ref}^i = \frac{1}{N}\sum_{t=1}^N(\phi_t^i \circ I_t^i)$ is calculated for groupwise registration, and each $I_t^i$ in image series $S^i$ should align its anatomical structures to $S_{ref}^i$. The $\mathcal{L}_\mathrm{similarity}$ term is defined as:
\begin{equation}
    \mathcal{L}_\mathrm{similarity} = - \frac{1}{N}\sum_{t=1}^N NMI(\phi_t^i \circ I_t^i, S_{ref}^i).
\end{equation}

In physics-informed test-time adaptation (PI-TTA), depending on the physics model of the test data, we generate the inversion recovery image series $\hat{S}^i$ following the relaxation model (using T1-relaxation as an example):
\begin{equation}
    \hat{S}^i(x, y, t_n) = |A(x, y) - B(x,y) \times e^{t_n/T_1^*(x, y)}|,
\end{equation}
where $A$, $B$, $T_1^*$ are the estimated parameters by fitting the warped images {$\phi_t^i \circ I_t^i$} to the 3-parameter model. The mutual information metric may handle the contrast variation but does not cope well with image content occlusion~\cite{xue2012motion}. With the inversion recovery image $\hat{S}^i$ following the same contrast pattern, the similarity term during fine-tuning is changed to mean squared error to capture pixel-wise differences, while the $\mathcal{L}_\mathrm{smooth}$, $\mathcal{L}_\mathrm{cyclic}$ remain the same as in pre-training. The pre-trained model is adapted for each individual test data.
\begin{figure}[ht!]
    \centering
    \includegraphics[width=\textwidth]{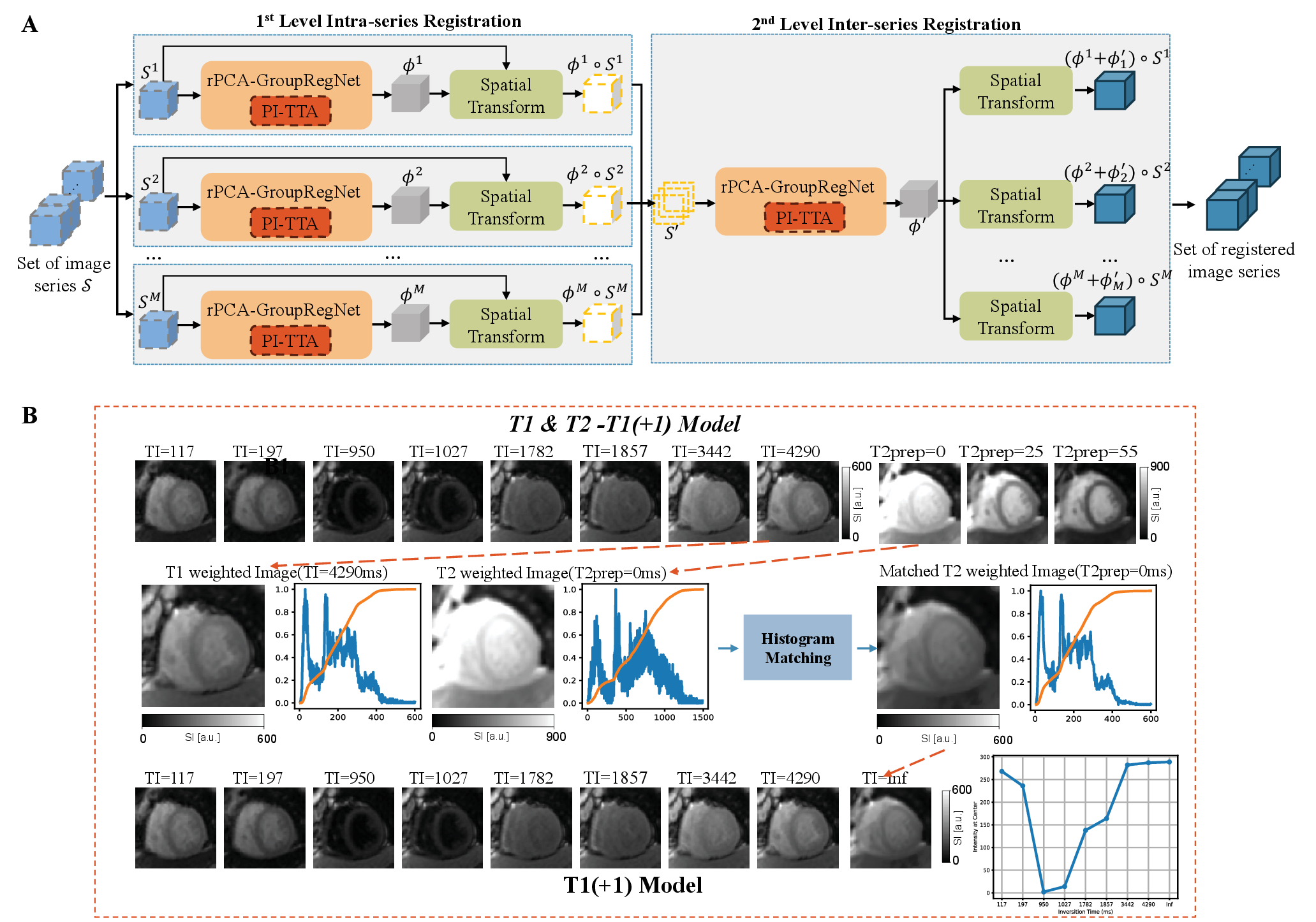}
    \caption{Hierarchical two-level registration pipeline. (A) The generalizable two-level registration pipeline is shown, where the moving volume is first registered within the sequences in the first level and then registered intra-sequence in the second level. (B) The physics model, T1+(1) model, applied in the second level for inter-subject registration are shown.}
    \label{fig:two-level}
\end{figure}

\subsection{Hierarchical Two-level Registration}
To register a set of image series that are described by multiple physics models (e.g., registration between T1 and T2 maps), we propose the hierarchical two-level registration pipeline as shown in \Cref{fig:two-level}. 

\noindent \textbf{Level 1 (Intra-acquisition registration):} Each image sequence $S^i$ is registered independently using the physics-informed test-time adaptation described in \Cref{sec:PITT}. This step corrects for motion within each individual acquisition sequence.

\noindent \textbf{Level 2 (Inter-acquisition registration):} The motion-corrected volumes from Level 1 are then registered across different acquisition protocols. The appropriate physical model is also applied based on the specific data pattern. To register the T1 and T2 weighted images, we apply the T1(+1) model as shown in \Cref{fig:two-level}. This approach treats the first T2-weighted image as a fully recovered T1-weighted image but in a different intensity range. The histogram matching is applied to normalize the first T2-weighted image to the same intensity scale as the last T1-weighted image. The combined series, including all T1-weighted images and the intensity-matched first T2-weighted image, then serves as input to the second-level registration using the T1(+1) physical model. The final deformation field is obtained by combining the transformations from both levels: $(\phi^i+ \phi^{\prime i})\circ S^i$, where $\phi^i$ represents the intra-acquisition deformation and $\phi^{\prime i}$ represents the corresponding inter-acquisition deformation.

\begin{figure}[ht!]
    \centering
    \includegraphics[width=0.95\textwidth]{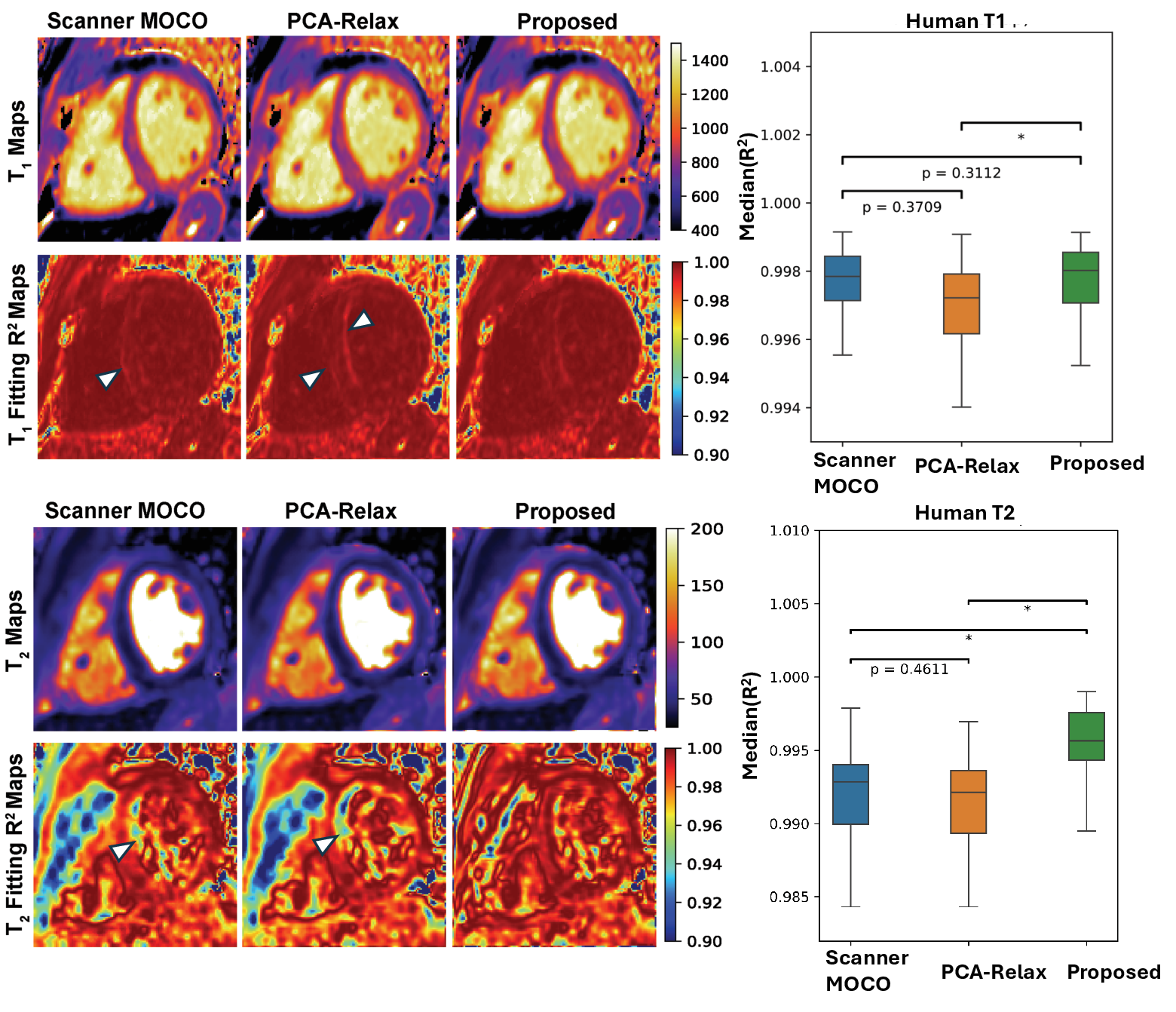}
    \caption{Representative figures and boxplots compare the performance of three methods on T1-weighted and T2-weighted data, using the scanner MOCO, pca-relax model and our proposed method. Our proposed approach (w/ PI) shows consistent improvement comparing to other approaches, especially around the myocardium boundary as indicated by the white arrows.}
    \label{fig:result_1}
\end{figure}
\section{Experiments and Results}
\subsection{Dataset and Implementation Details}
For model pre-training, we employed a cardiac MRI dataset comprising 48 subjects with post-contrast MOLLI sequences (Philips 3.0T) as our training data. Imaging was performed at three cardiac levels (base, mid-ventricular, and apex), with each subject contributing 1 to 3 slices. In total, 120 post-contrast MOLLI sequences were included. All images were resampled to a $224\times224\times11$ grid with 1 mm$^3$ isotropic resolution and followed by center cropping to $112\times112\times11$. For evaluation, we utilized a T1-T2 test dataset including 20 healthy volunteers, and each subject consisted of a pair of native T1 and T2 maps of the mid ventricular slice acquired during multiple breath holds.

The pre-trained model was trained using 50,000 steps, and fine-tuning for 10 steps for each test data. The rPCA-GroupRegNet~\cite{li2023contrast} was a convolutional neural network architecture based on the UNet~\cite{ronneberger2015u} architecture consisting of 4 encoding and 4 decoding layers with skip connections. Both encoder and decoder used convolutional blocks consisting of a 2D convolution and a Leaky ReLU activation function. And the rPCA (robust The number of time frames was considered as the batch information, thus, the batch normalization was used in the convolutional blocks. The smooth ($\lambda_0$) and cyclic ($\lambda_1$) regularization weight was set to 0.001 and 0.005 empirically. The curve fitting was calculated based on Levenberg-Marquardt minimization with parallel computing to accelerate the process.

\begin{figure}[h!]
    \centering
    \includegraphics[width=\textwidth]{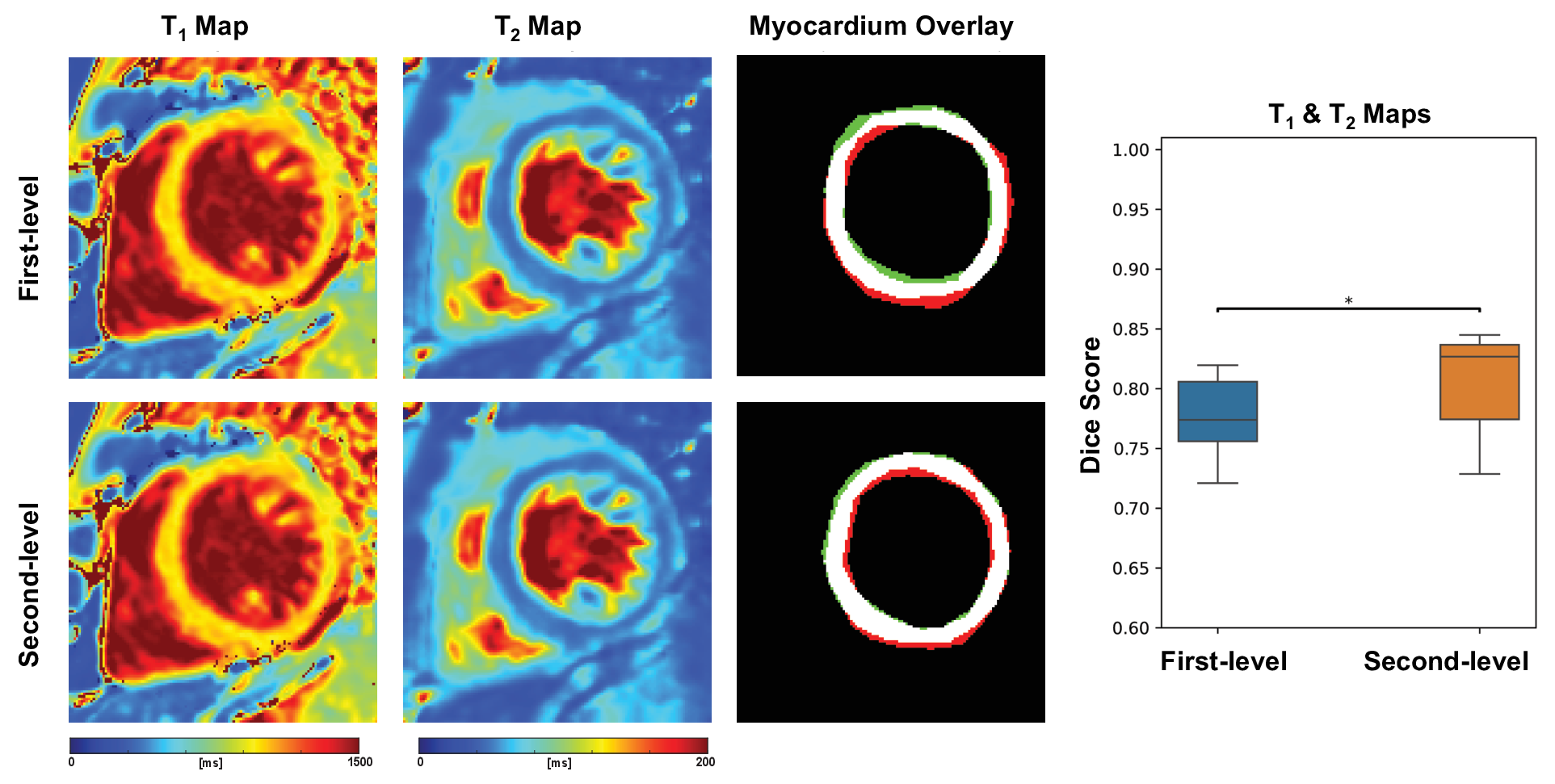}
    \caption{Representative figures and boxplots compare the performance between single-level and second-level registration. The representative figure show the improvement of the alignment between T1 and T2 maps using second-level registration. The red and green masks denote the T1 and T2 mapping's myocardial masks and the white region denote the overlay region. The statistical results showed significant improvement (p < 0.05) in dice score using second-level registration.}
    \label{fig:result_2}
\end{figure}
\subsection{Results}
Performance of the single-level registration was assessed by examining the goodness of fit from T1 and T2 maps within the myocardium. The model was compared with the scanner MOCO~\cite{xue2012motion}, pca-relax baseline model~\cite{zhang2024deep}, and our proposed model. Representative T1 and T2 maps and $R^2$ score maps (\Cref{fig:result_1}) demonstrated that the physics-informed test-time adaptation improved registration quality, particularly at the endocardial boundary. \Cref{fig:result_1} showed the statistical results measuring the fitting performance $R^2$ within the myocardium. The median $R^2$ showed a significant improvement in registration performance from the proposed method compared to prior art.

The effect of group registration across sequences was illustrated in \Cref{fig:result_2}, which showed the representative images and statistical results measuring the dice score~\cite{huizinga2016pca} within ROI (myocardium) between T1 and T2 maps, demonstrating a significant improvement in multiparametric CMR registration using the second-level registration.

\section{Discussion and Conclusion}

Our study highlights the improvements in image co-registration and motion correction facilitated by incorporating physics-informed test-time adaptation and hierarchical design into the registration pipeline. The physics-informed approach improves model performance on images with new contrast modulations that were not seen in the pre-trained model. The PI-TTA reduced the process to around 10 seconds comparing to the 2 minutes for one-shot approach, showing the feasibility for real-time clinical application. This flexible pipeline can be adapted for different multiparametric mapping techniques that are associated with multiple physical models and can serve as a tool for improving pixel-wise multiparametric mapping analysis. The results show the feasibility of using this approach on both T1 and T2 images on multiple human datasets. 

%
%
%
%
\newpage
\bibliographystyle{splncs04}
\bibliography{ref}
\end{document}